\documentclass[epj,10pt]{svjour}
\usepackage{amsfonts}
\usepackage{color}
\usepackage{eepic}
\usepackage{epic}
\usepackage{ulem}
\usepackage{graphicx}
\usepackage{texdraw}
\usepackage{epsfig}
\newcommand{\muMS}{\mu_{\bar{MS}}}
\begin{document}
\title{On the chromoelectric permittivity  and Debye screening in  hot QCD} 
\author{Vinod Chandra\inst{1} \thanks{E-mail:\email{ vinodc@iitk.ac.in}},
Akhilesh Ranjan \inst{1} \thanks{E-mail:\email{akranjan@iitk.ac.in}},
 V. Ravishankar\inst{1}\inst{2} 
\thanks{E-mail:\email{ vravi@iitk.ac.in, vravi@rri.res.in}}}
\institute{\inst{1} Department of Physics, Indian Institute of Technology 
Kanpur, UP, India, 208 016\\
\inst{2} Raman Research Institute, C V Raman Avenue, Sadashivanagar, 
Bangalore, 560 080, India}
\date{\today}
\abstract{
We study the response functions (chromo-electric susceptibilities)
for an interacting  quark-gluon plasma. The interaction effects have been  encoded in the effective fugacities
for quasi-partons which are extracted self-consistently from the two equations of state for hot QCD. The first one is  the fully perturbative  $O(g^5)$ EOS and, the second one which is  $O(g^6\ln(1/g))$, incorporates some non-perturbative effects.   We find that response function shows large deviations from the ideal behavior. We further determine the temperature dependence of the 
Debye mass by fixing the effective coupling constant $Q^2$ which appears in the transport equation. We show that our formalism naturally yields the leading order HTL expression for the Debye mass if we employ the ideal EOS. Employing the Debye mass, we estimate the dissociation temperatures for various charmonium and bottomonium bound states. These results are consistent with the
current theoretical studies.}

\maketitle


\section{Introduction}  
It is expected that at high temperatures ($T \sim 150-200 MeV$) and high densities ($\rho \sim 10 GeV/fm^{3}$) nuclear matter undergoes a deconfinement  transition to the quark-gluonic phase. 
This phase is under intense investigation in heavy ion collisions, and already, interesting results
have been reported by  Relativistic Heavy Ion Collider(RHIC) experiments \cite{expt}. As an important development, flow measurements\cite{star-report} suggest that close to the transition temperature $T_c$, the quark-gluon plasma (QGP) phase is strongly interacting ---
showing an almost perfect liquid behavior, with very low viscosity to entropy ratio --- rather than  showing a behavior close to that of an ideal gas. See ref. \cite{new-matter} for a 
  comprehensive review of experimental observations from RHIC, and   ref. \cite{expt,exp-status1,exp-status2,s-a-bass,csorgo} for other recent
experimental results. On the other hand,
lattice computations \cite{lattice,lattice-new} also suggest that QGP is strongly interacting even at $T=2T_c$. This finding has been reproduced by a number of other theoretical studies ---
 by employing AdS/CFT correspondence in the strongly interacting regime of QCD\cite{dtson},  by molecular dynamical simulations for classical strongly coupled systems\cite{shuryak}, and  by  model calculations with Au-Au data from RHIC \cite{bair,jyo}.
  In the backdrop of the above developments, a number of standard diagnostics, such as $J/\psi$ suppression and strangeness enhancement, which have been proposed to probe QGP also need to be re-examined. It is also of importance to address other transport properties, production and equilibration dynamics, and the physical manifestations of pre-equilibrium evolution.

If  this is the case, as it indeed appears to be, then the plasma interactions would be largely in the non-perturbative regime; in this regime,  few analytic techniques  are available for a robust theoretical analysis. Effective interaction approaches are needed.
 In  this direction, considerable work has already been done and we refer the reader to ref. \cite{matsui,elze,vr1,vr2,akranjan,shur2,shur3,blaizot1} for some of the theoretical results.

The effective approaches emphasize the collective origin of the plasma properties which can be best
understood within a semi-classical framework. Indeed, in a recent work
\cite{fluid}, the  successes of  hydrodynamics in interpreting and understanding the experimental observations from RHIC has been reviewed.  Since more exciting and discerning data are expected from LHC experiments soon, and given the above context, it is worthwhile
exploring semi-classical techniques to understand the properties of QGP in heavy ion collisions.
In this context, it is known by
now \cite{pisarski,braaten-1,braaten-2,nair} that a classical behavior emerges 
naturally when one considers hard thermal loop(HTL) contributions. A local 
formulation of HTL effective action has been obtained by Blaizot and Iancu who
have succeeded in rewriting the HTL effective theory as a kinetic theory with
a Vlasov term \cite{blaizot,blaizot-2,blaizot-3,blaizot-4}. A significant development in this direction
is the realization that the HTL effects are, in fact, essentially classical and that they are much
easier to handle within the frame work of classical transport equations \cite{cm1,cm4,cristina}.  Thus,
the semi-classical techniques appear to hold the promise of providing tools to understand the bulk properties of QGP.

The present paper continues the theme, and its
 central aim  is to combine the kinetic equation approach which yields the transport properties, with the hot QCD equations of state to make predictions which can  perhaps be tested in heavy ion collisions. Recently, Ranjan and Ravishankar have developed a systematic approach to determine  fully the response functions  of QGP, with a special emphasis on the color charge as a dynamical variable \cite{akranjan}.
 In parallel,
Chandra, Kumar and Ravishankar have succeeded in adapting  two hot QCD EOS to make predictions for heavy ion collisions \cite{chandra1,chandra3}. They have shown that  the 
interaction effects which modify  the equations 
of state can be expressed by 
 absorbing them into effective fugacities ($z_{q,g}$) of otherwise free or weakly interacting quasi quarks and gluons. 
 Since the analysis in ref. (\cite{akranjan}) was illustrated only for (the academically interesting) case of ideal quarks and gluons, it is but natural to bring the two studies together and explore what the hot QCD EOS have to predict for heavy ion collisions.
We take up this program in this paper. 

We determine the Debye mass as a function of temperature first by combining the quasi-particle description of improved pQCD EOS  with the semi-classical transport theory formulation of response functions for QGP. Employing the Debye mass so determined, we have estimated the dissociation temperatures for various quarkonia states.
For our purpose, we consider two specific  hot QCD equations of state: The first, which we call EOS1 is
perturbative, with contributions up to 
$O(g^5)$\cite{arnold,zhai}. The second EOS has  a free parameter $\delta$, and is evaluated
upto $O[g^6\log(1/g)]$\cite{kaj1}. We denote it by EOS$\delta$.  $\delta$ may be
fine tuned to get a reasonably good agreement \cite{kaj1} with the lattice results
\cite{fkarsch}, which we exploit here.  Both the EOS are expected to be valid for $T \ge 2 T_c$ \cite{kaj1}.

 The paper is organized as follows:  In section 2, we  introduce  the two
hot QCD equations of state and outline the recently developed 
method\cite{chandra1} to adapt them for making definite predictions for QGP at RHIC 
and the forthcoming experiments at LHC. In section 3, we  obtain the expressions for 
the response functions of interacting QGP and in section 4, we  study
their temperature dependence in detail. Moreover, we  study 
the Debye screening and the dissociation phenomenon of  heavy quarkonia 
states in hot QCD medium.
In doing so we also relate the phenomenological charge that occurs in the transport
equation to lattice and experimental observables. We conclude the paper in section 5.

\section{Hot QCD equations of state and their quasi-particle description}
There are various equations of state proposed for QGP at RHIC. These include 
non-perturbative lattice EOS \cite{fkarsch}, hard thermal loop(HTL) resumed 
EOS\cite{htleos} and perturbative hot QCD equations of state
\cite{zhai,arnold,kaj1}. In the present paper, we seek to determine the 
chromo-electric response functions for QGP by employing two EOS:
(i) the fully perturbative $O(g^5)$ hot QCD EOS 
proposed by Arnold and Zhai\cite{arnold} and Zhai and Kastening
\cite{zhai}, and (ii) The EOS of   $O[g^6(\ln(1/g)+\delta)]$  determined by Kajantie {\it et al}
\cite{kaj1}, by incorporating contributions from non-perturbative 
scales, $gT$ and $g^2T$. We employ the method
recently formulated by Ranjan and 
Ravishankar\cite{akranjan} to extract the chromo-electric permittivities of the medium.  
The  EOS which we label EOS$\delta$ is given by
\begin{eqnarray}
\label{eq1}
P_{g^6\ln(1/g)}&=&\frac{8\pi^2}{45\beta^4}\bigg \lbrace (1+\frac{21N_f}{32})-\frac{15}{4}(1+\frac{5N_f}{12})\frac{\alpha_s}{\pi}
\nonumber\\ &&+30(1+\frac{N_f}{6})(\frac{\alpha_s}{\pi})^{\frac{3}{2}} 
+\bigg[237.2+15.97N_f\nonumber\\ 
&&-0.413 N_f^2 +\frac{135}{2}(1+\frac{N_f}{6})\ln(\frac{\alpha_s}{\pi}(1+\frac{N_f}{6}))\nonumber\\
&&-\frac{165}{8}(1+\frac{5N_f}{12})(1-\frac{2N_f}{33})\ln[\frac{\muMS\beta}{2\pi}]\bigg](\frac{\alpha_s}{\pi})^2\nonumber\\
&&+(1+\frac{N_f}{6})^{\frac{1}{2}}\bigg[-799.2-21.99N_f-1.926N_f^2\nonumber\\
&&+\frac{495}{2}(1+\frac{N_f}{6})(1+\frac{2N_f}{33})\ln[\frac{\muMS\beta}{2\pi}]\bigg](\frac{\alpha_s}{\pi})^{\frac{5}{2}} 
 \nonumber\\
&&+\frac{8\pi^2}{45}T^4 \biggl[1134.8+65.89 N_f+7.653 N_f^2\nonumber\\
 &&-\frac{1485}{2}\left(1+\frac{1}{6} N_f\right)\left(1-\frac{2}{33}N_f\right)
\ln(\frac{\muMS}{2\pi T})\biggr]\nonumber\\
&&\times\left(\frac{\alpha_s}{\pi}\right)^{3}
(\ln \frac{1}{\alpha_s}+\delta)\bigg \rbrace.\nonumber\\
\end{eqnarray}
EOS1 is obtained from this equation by dropping the last term which has contributions of $O(g^6(\ln(1/g)+\delta))$. The phenomenological parameter $\delta$ is introduced in \cite{kaj1} to incorporate the undetermined contributions
of $O(g^6)$. 
It also acts as  a fitting parameter to get the best
agreement with the lattice results. 
\begin{figure}[htb]
\label{fig1}
\vspace*{-78mm}
\hspace*{-45mm}
\psfig{figure=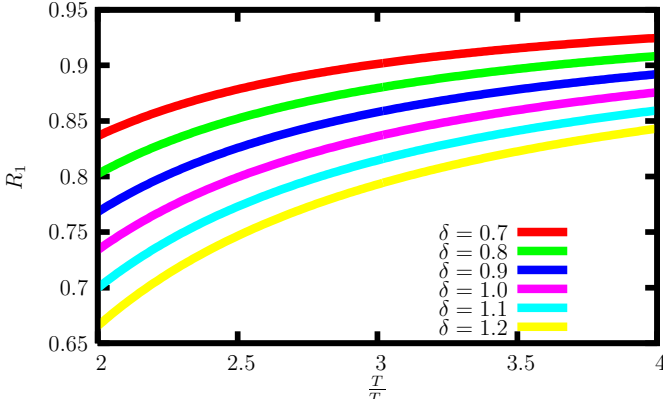,width=150mm}
\vspace*{-90mm}
\caption{(Color online) Relative equation of state  {\it wrt} ideal EOS for pure gauge theory plasma 
as a function of $T/T_c$ for various values of $\delta$.}
\end{figure}.

\begin{figure}[htb]
\label{fig2}
\vspace*{-78mm}
\hspace*{-50mm}
\psfig{figure=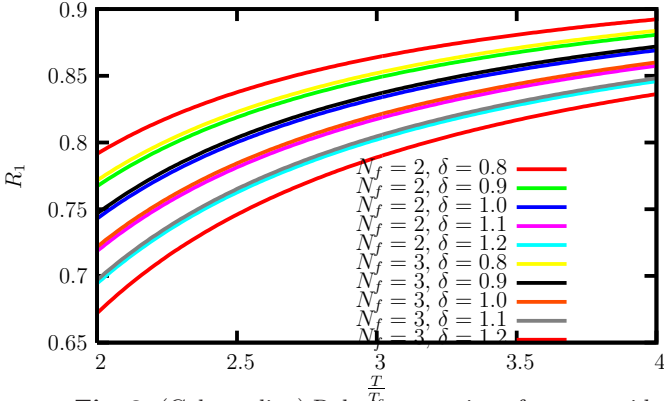,width=150mm}
\vspace*{-90mm}
\caption{(Color online) Relative equation of state {\it wrt} ideal EOS for full QCD plasma 
with $N_f=2,3$ as a function of $T/T_c$  for various values of $\delta$.} 
\end{figure}

EOS1 and EOS$\delta$ have several ambiguities, associated with the renormalization scale
$(\mu_{\bar MS})$, the scale parameter $\Lambda_T/\Lambda_{\bar{MS}}$ which occurs in the expression for the running coupling constant $\alpha_s$, and the value of the phenomenological parameter $\delta$.
The arbitrariness in fixing $(\mu_{\bar MS})$ has been discussed well in literature and a popular way out is the BLM criterion due to Brodsky, Lepage and Mackenzie \cite{blm}. In this criterion, the value of $(\mu_{\bar MS})$ is allowed to  vary between $\pi T$ and 4$\pi T$ \cite{neito}. Here, we choose $\mu_{\bar MS}=2.15\pi T\approx 6.752 T$\cite{avrn1} close to the central value $2\pi T$.
With this  particular choice,  all the contributions due to the logarithms containing $\mu_{\bar MS} $ are very small.
 For the scale parameter  $\Lambda_T$, we follow Huang and Lissia \cite{shaung} and set $\Lambda_T/\Lambda_{\bar MS}=\exp(\gamma_E + 1/22)/4\pi \approx 0.148$. They found that this choice makes the coupling $g^2(T)$ is optimal for lattice perturbative calculations. The same value has also been employed by others. see e.g. \cite{kaj1,avrn1}.

\subsection{The underlying distribution functions}
The  construction of  the distribution functions that underlie the
EOS, in terms of effective quarks and gluons which act as quasi-excitations, has been discussed by Chandra {\it et. al}\cite{chandra1} in the specific context of EOS1 and EOS$\delta$. To review the method briefly, {\sl all} the terms that represent interactions are collected together by recasting them as effective fugacities ($z_{q,g}$) for the otherwise free quarks and gluons. Of course, the pure gauge theory
case is simply obtained by putting the number of flavors, $N_f=0$ in the EOS. Thus, $z_g$ represents the self interactions of the gluons, while $z_q$ encapsulates the quark-quark and the quark gluon interaction terms. The quantities $z_{g/q}$ can be determined 
from the two EOS self-consistently. In this procedure,  all the temperature
effects are contained in the effective fugacities  $z\equiv z(\alpha_s(T/T_c))$, where we display the dependence on the temperature and coupling constant  explicitly. It has been shown in ref. \cite{chandra1} (where the details can be found) that one can trade off the dependence of the effective fugacities on the renormalization scale ( $\mu_{\bar{MS}}$) by their dependence on the critical temperature $T_c$. For that purpose, one utilizes the one loop expression of $\alpha_s(T)$ at finite temperature given by \cite{shaung}.
It should be borne in mind  that the effective fugacity $z_{g/q}$ introduced here has nothing to do with the 
usual fugacity which corresponds to the conservation of particle number. It is, therefore, unrelated to the baryon chemical potential in the case of quarks. $z_{g/q}$ merely encodes
  the interaction effects present in hot QCD EOS. Its deviation from unity signifies the presence of non-ideal terms in the EOS.

In a recent work Chandra and Ravishankar\cite{chandra3} showed that the 
effective fugacities $z_{g/q}$ modifies the dispersion relations. The modification captures 
the non-zero trace anomaly effects in hot QCD. They further employed this model to 
determine the shear viscosity and shear viscosity to entropy ratio as a function of temperature. This model has further been generalized to the pure lattice gauge theory 
EOS in \cite{chandra4} and the authors found that the model works remarkably well for lattice QCD. This quasi-particle description is analogous to  Landau's theory of Fermi liquids. The authors\cite{chandra5} have further determined the temperature dependence of gluon quenching parameter and studied
the shear viscosity ($\eta$), and its ratio with entropy ($\eta/{\mathcal S}$) 
employing the recent work by Asakawa, M\"uller and Bass \cite{asakawa} and  
Majumder,  M\"uller and  Wang\cite{majumder}.

In  figs. 1 and 2,  we display the behavior of
EOS$\delta$ for various values of the parameter $\delta$. The figures show the  pure gauge theory contributions to the EOS and full QCD separately. We remark parenthetically that the studies in the earlier  work \cite{chandra1} were confined to EOS1 and the special case $\delta=0$ in EOS$\delta$. For the details on EOS1 and EOS$\delta$ for $\delta=0$, we refer the reader to ref. \cite{chandra1} (see fig.1-7 of ref.\cite{chandra1}).
 First of all, we see that as $\delta$ increases in magnitude, the EOS, for both pure gauge theory and full
QCD, become softer, with $P/P_I$ taking smaller values, we denote the the ratio $P/P_I$ by $R_1$.
Kajantie\cite{kaj1} 
obtains  the best fit with the lattice results of Boyd et. al.\cite{boyd} 
by choosing a value $\delta =0.7$. We find  that to get agreement with the
more recent results of Karsch \cite{fkarsch}, $\delta \approx 1.0$ is preferred, when we consider $T >2T_c$. Over all, we find that the range of values $0.8 \le \delta \le 1.2$ gives a reasonably good qualitative agreement with the lattice results for the screening lengths. Finally, we set   $\Lambda_{\bar{MS}} = T_c$, which is close to the value  $0.87T_c$ found by Gupta \cite{sgupta}.
Here, we wish to mention that there is an uncertainty in fixing the free parameter $\delta$. This follows from the 
freedom in choosing the  QCD renormalization scale at high temperature. This has been investigated in detail by Blaizot, Iancu and Rebhan \cite{rebh}.
 The value of $\delta$ in the present paper has been obtained by employing the one loop expression for the running coupling constant and the QCD renormalization scale determined  in ref.\cite{shaung}. 

The behavior of the corresponding fugacities, as a function of temperature, is shown in fig.3. It may be seen that $0<z_{g,q}<1.0$ which ensures the convergence of the method to determine the effective fugacities from the hot QCD EOS. 
We now proceed to determine the response of the plasma in the next section.

\begin{figure}[htb]
\label{fig3}
\vspace*{-78mm}
\hspace*{-45mm}
\psfig{figure=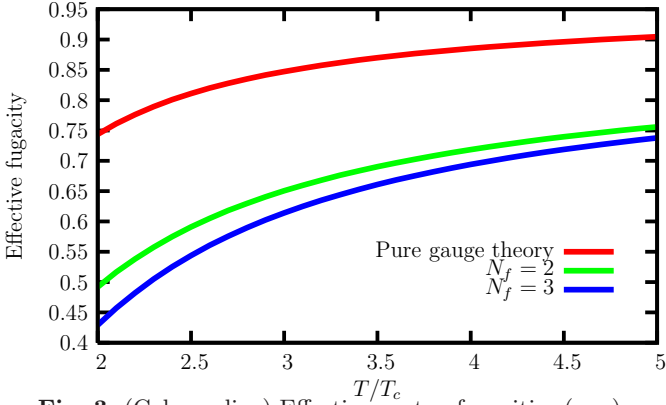,width=150mm}
\vspace*{-90mm}
\caption{(Color online) Effective parton fugacities $(z_{g,q})$ quarks determined from EOS$\delta$ as a function of temperature. Note that the behavior is shown for  $\delta=1.0$.
}
\end{figure}.

\section{Response functions for interacting qgp}
Recently Ranjan and Ravishankar \cite{akranjan} have determined the form of 
chromo-electric response functions for collision less quark-gluon plasma within 
the framework 
of semi-classical transport theory. They have set up the transport equation in 
the extended phase space including the SU(3) group space corresponding to 
dynamical color degree of freedom. They have taken the distribution function
in a coherent state basis defined over the extended single particle phase space 
$\mathcal{R}^6 \otimes \mathcal{C}_{\mathcal{G}}$, where 
$\mathcal{C}_{\mathcal{G}} = \mathcal{G} /\mathcal{H}$ is the phase space 
corresponding to the color degree of freedom, obtained as a coset space by 
factoring the group space by the stabilizer group $\mathcal{H}$ of any 
reference state in the Hilbert space. Having been employed to study the ideal 
case, the formalism has not been applied to examine the behavior of the 
plasma with a realistic EOS. We employ the results of the previous section and 
rectify this drawback, by incorporating the interaction effects as represented 
by EOS1 and EOS$\delta$.

A brief comment on the response functions. In contrast to electrodynamic plasma, 
the chromo-electric response has a richer structure. Apart from the standard 
permittivity which we shall call Abelian and denote by $\epsilon_A$, there are 
additional response functions, their number depending on the color carried by 
the partons. Thus, quarks have an additional response function which affects the 
non-Abelian coupling. The corresponding permittivity will be called non-Abelian, 
and denoted by $\epsilon_N$. The two functions exhaust the response in the quark 
sector. The gluonic sector, arising from the adjoint representation of the gauge 
group admits yet another kind of response, corresponding to tensor excitations. 
These excitations are not allowed in the quark sector (which emerges from the 
fundamental representation of the gauge group).
We consider each of these response functions for the interacting QGP. The 
response functions are obtained in the temporal gauge. 

Consider first the familiar Abelian component of the response $\epsilon_A$. For 
an isotropic plasma(in the absence of chromo-magnetic fields), its expression is 
given by \cite{akranjan}
\begin{eqnarray}
\label{eqn3}
\tilde{\epsilon}_A(\omega,\vec{k})=1+Q^2I_0(\omega,\vec{k})
\end{eqnarray}
where $Q^2 = Q^aQ^a$ is the color charge magnitude squared, and $I_0$ is 
determined by the equilibrium distribution function thus:
\[ \int \frac{1}{\omega-\frac{\vec{k}\cdot\vec{p}}{\varepsilon}+i\tau}\frac{\partial f_{eq}}{\partial p_i} \,d^3\vec{p}\equiv k_i I_0 (\omega,\vec{k}),\] 

The non-Abelian response function, which has been evaluated in the long 
wavelength limit, is given by
\begin{eqnarray}
\label{eqn4}
\tilde{\epsilon}_N(\omega,\omega')=\{1+\frac{Q^2\left.I_1(\omega',\vec{k}')\right|_{\vec{k}'=0}}{\omega}\}
\end{eqnarray}
where $I_1$ is defined as
\[I_1(\omega,\vec{k})=\frac{1}{3}Tr\left(\int \frac{\frac{p_j}{\varepsilon}}{(\omega-\frac{\vec{k}\cdot\vec{p}}{\varepsilon})}\frac{\partial f_{eq}}{\partial p_i} \,d^3\vec{p}\right).\]
We recall that the new constitutive Yang-Mills equations, in the presence of 
the medium, are given by

\begin{eqnarray}
\label{eqn5}
\tilde{\rho}^a(\omega,\vec{k})+iQ^2\tilde{E}^a_i(\omega,\vec{k})k_iI_0(\omega,\vec{k})\nonumber \\
-\frac{Q^2f^{alm}}{\omega} \int \left.I_1(\omega',\vec{k}')\right|_{\vec{k}'=0} \tilde{A}^l_i(\omega-\omega',\vec{k}-\vec{k}')\nonumber \\
\times \tilde{E}^m_i(\omega',\vec{k}') \,d\omega' \, d^3\vec{k}'=0.
\end{eqnarray}

\begin{eqnarray}
\label{eqn6}
\tilde{j}^a_j(\omega,\vec{k})+iQ^2\tilde{E}^a_i(\omega,\vec{k})\delta_{ij}\left.I_1(\omega,\vec{k})\right|_{\vec{k}=0}\nonumber \\
=0.
\end{eqnarray}

As pointed out in \cite{akranjan}, the Abelian and non-Abelian responses are not 
independent of each other. Gauge invariance relates them, by virtue of which we 
can obtain both from a common generating function as follows:

\begin{eqnarray}
\label{eqn7}
I_0= \frac{1}{k^2}\frac{\partial}{\partial \omega} \int ln(\omega -
\frac{\vec{k}\cdot \vec{p}}{\varepsilon})k_i \partial_{{p_i}}f_{eq} d^3p \nonumber \\
I_1= -\frac{1}{3}Tr\bigg(\frac{\partial}{\partial k_j} \int ln(\omega -
\frac{\vec{k}\cdot \vec{p}}{\varepsilon}) \partial_{p_i}f_{eq} d^3p\bigg).
\end{eqnarray}

We further recall that these expansions are determined when the system is 
displaced slightly from its equilibrium, in the collisionless limit.

\subsection{Ideal response}
It is convenient to first write the expressions for the responses of ideal 
distributions for quarks and gluons. The responses due to EOS1 and EOS$\delta$ 
get a simple modification over their ideal forms since we have mapped 
successfully the interaction effects into quasi free partons with effective 
fugacities. Thus, in the ideal case we have, for the quarks,

\begin{eqnarray}
\label{eqn8}
\tilde{\epsilon}_A^{(q)}=[1+\frac{2\pi^3Q^2T^2N_f}{3k^2}\{-\frac{\omega}{k}ln\left|\frac{\omega+k}{\omega-k}\right|+2\}]\nonumber\\
\end{eqnarray}

and the non-Abelian response function is given by
\begin{eqnarray}
\label{eqn9}
\tilde{\epsilon}_N^{(q)}
=\{1-\frac{4\pi^3 Q^2T^2N_f}{9}\frac{1}{\omega\omega'}\}.
\end{eqnarray}

The imaginary part of Abelian($\tilde\epsilon_A$) and non-Abelian component ($\tilde\epsilon_N$) of the chromo-electric permittivity can be easily evaluated by the standard Landau $i\epsilon$ prescription. These are needed to obtain landau damping which we do not study here.
The gluon effective charge($Q^2_g$) and the quark effective charges ($Q^2_q$) 
are related to each other by the total number of color $N_c$ for 
$SU(N_c)$ as $Q^2_g=N_c Q^2_q\equiv N_c Q^2$, where Q is the color charge of a single  quark. 
It is easy to see that the 
gluonic susceptibility, $\chi^{(q)}\equiv \tilde{\epsilon}_{A,N}^{g}-1$ is identical to the quark susceptibility, $\chi^{(q)}\equiv \tilde{\epsilon}_{A,N}^{q}-1$; they are related as,
\begin{equation}
\chi^{(q)}=\frac{N_f}{2N_c} \chi^{(g)}
\end{equation}

\subsection{Interaction effects}
We now consider the modification that the above expressions undergo permittivities arising because of the new EOS. Recall 
that the corresponding equilibrium distribution functions differ from each 
other only in their form for the chemical potentials $\mu_{q,g}$. The responses 
thus depend on the interactions implicitly through an explicit dependence 
on $z_{q,g}$.

 Considering the gluonic case, {\sl i. e.}, pure gauge theory first, we 
get the expressions for the two permittivities as

\begin{eqnarray}
\label{eqn10}
\tilde{\epsilon}_A=[1+\frac{4\pi^3N_c Q^2T^2 g_2^{\prime}(z_g)}{3k^2}\{-\frac{\omega}{k}ln\left|\frac{\omega+k}{\omega-k}\right|+2\}],
\end{eqnarray}

 and the non-Abelian response function as

\begin{eqnarray}
\label{eqn11}
\tilde{\epsilon}_N
=\{1-\frac{8\pi^3 N_c Q^2T^2 g_2^{\prime}(z_g)}{9}\frac{1}{\omega\omega'}\}.
\end{eqnarray}
The function  $g_2^{\prime}(z_g) \equiv \frac{6}{\pi^2} g_2(z_g)$ where 
$g_2(z_g)$ is defined via the integral  below.

$$
\int_0^{\infty} \frac{x^{\nu-1}}{z_g^{-1}exp({x})-1}\,dx=\Gamma(\nu)g_{\nu}(z_g).\
$$
$g_{\nu}(z_g)$ has the series expansion
$$
g_{\nu}(z_g)=\sum_{l=1}^{\infty} \frac{z_g^l}{l^{\nu}} ~~\mbox{for}~~ z_g\ll 1.
$$
Note that $g'_2(1)=1$ gives the ideal limit.

Similarly, the corresponding expressions in the quark sector are obtained as

\begin{eqnarray}
\label{eqn12}
\tilde{\epsilon}_A=[1+\frac{2\pi^3Q^2T^2N_f f^{\prime}_2(z_q)}{3k^2}\{-\frac{\omega}{k}ln\left|\frac{\omega+k}{\omega-k}\right|+2\}]\nonumber\\
\end{eqnarray}

and the non-Abelian response for effective quarks reads:

\begin{eqnarray}
\label{eqn13}
\tilde{\epsilon}_N
=\{1-\frac{4\pi^3 Q^2T^2N_f f^{\prime}_2(z_q)}{9}\frac{1}{\omega\omega'}\}.
\end{eqnarray}

The function  $f_2^{\prime}(z_q) \equiv \frac{12}{\pi^2} f_2(z_q)$ where 
$f_2(z_q)$ is defined via the integral below.

\[\int_0^{\infty} \frac{x^{\nu-1}}{z_q^{-1}exp({x})+1}\,dx=\Gamma(\nu)f_{\nu}(z_q)\]
\[f_{\nu}(z_q)=\sum_{l=1}^{\infty} (-1)^{l-1} \frac{z_q^l}{l^{\nu}} ~~\mbox{for}~~ z_q\ll 1\]
and $f'_2(1)=1$.
 
\section{Effective charges and relative susceptibilities}
eq.(\ref{eqn10}-\ref{eqn13}) admit a simple physical interpretation, when compared with their
counterparts eq.(\ref{eqn8} -\ref{eqn9}). Indeed, the sole effect of the interactions on the transport properties is to merely renormalize the   
the quark and the gluon charges $Q_{g,q}$ as shown 
below:
$$
Q_g^2 \rightarrow \bar{Q}_g^2 = Q_g^2 g_2^{\prime}(z_g); ~~Q_q^2 \rightarrow \bar{Q}_q^2 = Q^2 f_2^{\prime}(z_q).
$$
The renormalization factors $g_2^{\prime}(z_g), f_2^{\prime}(z_q)$ further possess the significance of
chromo-electric susceptibilities, relative to the ideal values. To see that, we 
note that the Abelian and the non-Abelian strengths for gluons as well as quarks 
suffer the same renormalization reflecting the underlying gauge invariance. 
Furthermore, the expressions for the relative susceptibilities are given by, 
\begin{eqnarray}
\label{eqn14}
{\cal R}=\frac{\chi(z)}{\chi(1)}\equiv \frac{{\cal A}(z)}{{\cal A}(1)}=\frac{{\cal N}(z)}{{\cal N}(1)}= \left\{ \begin{array}{rcl}
f'_2(z_q) &\mbox{for quarks,} &\\
g'_2(z_g) &\mbox{for gluons}&
\end{array} \right.
\end{eqnarray}
and
\begin{eqnarray}
\label{eqn15}
{\cal R}_{q,g}=\frac{\chi^{(q)}(z_q)}{\chi^{(g)}(z_g)}\equiv\frac{{\cal A}^{(q)}(z_q)}{{\cal A}^{(g)}(z_g)}=\frac{{\cal N}^{(q)}(z_q)}{{\cal N}^{(g)}(z_g)}=\frac{N_f}{2 N_c}\frac{f'_2(z_q)}{g'_2(z_g)}.
\end{eqnarray}
Note that the relative susceptibilities are entirely functions of the single variable $T/T_c$, and are
independent of ($\omega, k$). The dependence of the susceptibilities on ($\omega, k$) has already been studied in
detail in ref.\cite{akranjan}. We merely concentrate on the temperature dependence below.

Before we go on to discuss the susceptibilities and other bulk properties, 
 we point out an essential care to be taken in using the above 
susceptibilities for determining the response of the plasma. For pure gauge theory, only 
the gluonic part contributes, while for the full QCD, we have to necessarily 
take the contribution from both the quark and the gluonic sector. We discuss 
both the cases below. The response functions for the full QCD is obtained by 
summing over the response function for quark as well as 
gluon plasma. The relative susceptibility for full QCD plasma is given by
\begin{eqnarray}
\label{eqn16}
{\cal R'}=\frac{N_f f'_2(z_q)+2N_cg'_2(z_g)}{N_f+2N_c}.
\end{eqnarray} 

\subsection{Behavior of the susceptibilities}
We now proceed to study the behavior of the relative susceptibilities displayed 
in eqs.(\ref{eqn14}), (\ref{eqn15}) and (\ref{eqn16}) as  functions of temperature.  As observed, relative susceptibilities for both quarks and 
gluons scale with $T/T_c$. 
We have plotted the relative susceptibilities ${\mathcal R}$, 
${\mathcal R_{qg}}$ and ${\mathcal R}^\prime$ as  functions of $T/T_c$
(See figs.4-6), for both EOS1 and EOS$\delta$.  Please note that 
we have chosen $\delta=1.0$ in EOS$\delta$.

%\subsubsection{Pure QCD -- the gluonic plasma}
Fig.4 shows the relative susceptibility of a purely gluonic plasma as a function of 
temperature for EOS1 and EOS$\delta$.

\begin{figure}[htb]
\label{fig4}
\vspace*{-70mm}
\hspace*{-45mm}
\psfig{figure=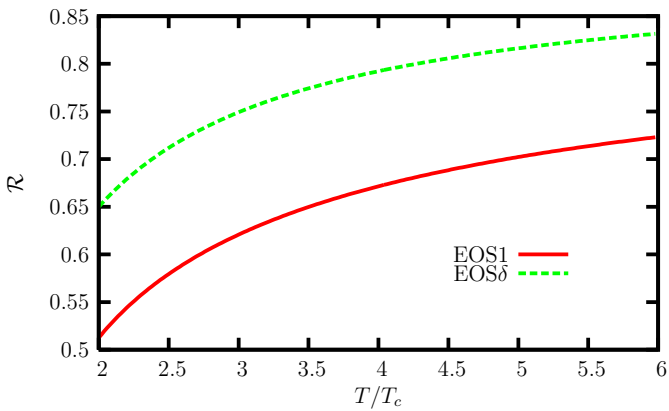,width=150mm}
\vspace*{-80mm}
\caption{(Color online) Relative susceptibility, $g'_2(z_g)$ (see eq. (\ref{eqn14}), for  pure gauge theory plasma 
 as a function of $T/T_c$ for EOS1 and EOS$\delta$ ($\delta=1$).}.  
\end{figure}

\begin{figure}[htb]
\label{fig5}
%\begin{center}
\vspace*{-80mm}
\hspace*{-43mm}
\psfig{figure=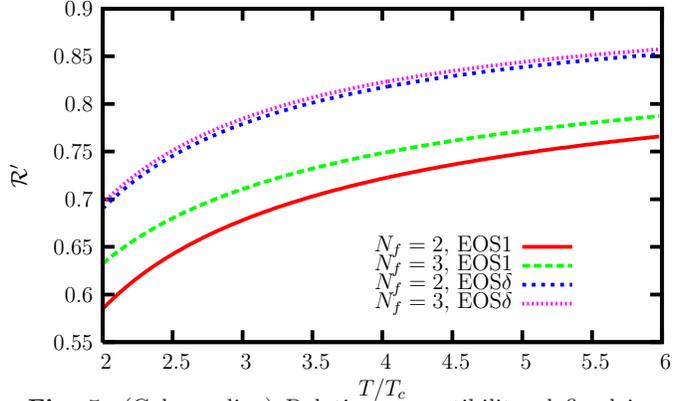,width=150mm}
\vspace*{-90mm}
\caption{(Color online) Relative susceptibility, defined in eq. (\ref{eqn16}), for the full QCD plasma 
 as a function of $T/T_c$, for EOS1 and EOS$\delta$ ($\delta=1)$. We have studied the cases $N_f= 2,3$.}
\end{figure}  
We see From fig. 4 that the susceptibility of a purely gluonic plasma is weaker in the presence of interactions, approaching its ideal value asymptotically
with increasing temperatures. Equivalently, there is a decrease in the value of the phenomenological coupling
$Q^2$, relative to its ideal value.

The behavior of quark gluon plasma is not qualitatively different from that of a purely gluonic plasma,
as may be seen from fig.5. In other words, the quark contribution is of the same order as the purely gluonic contribution. However, the relative contribution from the quarks and the gluons does depend on the EOS considered. Indeed, with EOS1 (where interactions up to $O(g^5)$ are included), fig.6 shows that the gluonic contribution dominates over the quark contribution for both $N_f=2$ and $N_f=3$ and  same is true for  EOS$\delta$. The dominance is more pronounced for EOS1 as comapare to EOS$\delta$. 

At this juncture, we clarify that 
 by ideal we mean that we employ the ideal EOS . The expressions so obtained  are the  same as the one loop result, or
equivalently,  the leading order HTL result. Similarly, the phrase interacting QGP refers to the contribution of the non-ideal terms in the hot QCD EOS,  which is equivalent to the inclusion of higher order  corrections to the one loop result.

\begin{figure}[htb]
\label{fig6}
\vspace*{-80mm}
\hspace*{-43mm}
\psfig{figure=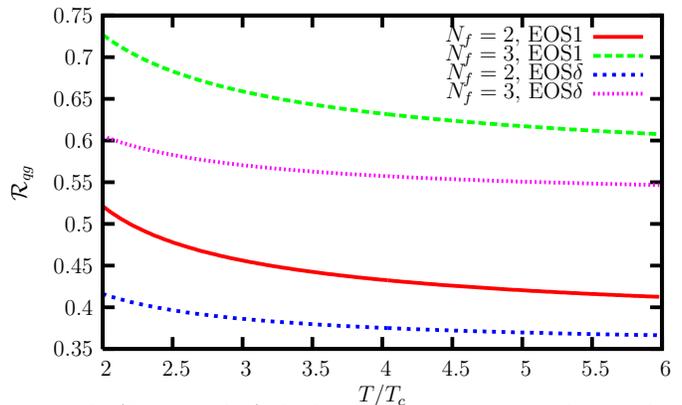,width=150mm}
\vspace*{-90mm}
\caption{(Color online) Ratio of the quark to gluonic contributions to the susceptibility (see eq.(\ref{eqn15}) 
as a function of $T/T_c$ for EOS1 and EOS$\delta$ for $\delta=1$
}
\end{figure}.

\subsection{Debye mass from the chromoelectric susceptibility}
The static or zero frequency limit of the response function 
leads to the well known relation with the Debye mass as follows,
$\epsilon(\vec k,T)=1+\frac{M^2_D}{k^2}$. 

One can compute the Debye mass for full QCD by utilizing the additive nature 
of Abelian response functions of partons in QCD.
The full permittivity for hot QCD plasma will be obtained, 
\begin{eqnarray}
\label{eqs}
\epsilon(k,T)&=&1+\chi^{(g)}(z_g) +\chi^{(q)}(z_q)
\end{eqnarray}
which naturally incorporates the additivity of Debye mass square in thermal QCD.

For an ideal QGP the Debye mass comes out to be,
\begin{equation}
(M^I_D)^2= 8\pi^3 Q^2 T^2 (N_c/3+N_f/6). 
\end{equation}
Interestingly, this matches with the leading order HTL expression for 
the Debye mass if one identifies the phenomenological charge Q as
\begin{equation}
 Q^2=\frac{g^2(T)}{8\pi^3},
\end{equation}
where $g(T)$ is the QCD running coupling constant at finite temperature.
This fixes the phenomenological charge Q. We employ this to study 
how the inclusion of intercations influence the Debye screening and the 
dissociation phenomenon of heavy quarkonia in a hot QCD medium.

 The current approach to determine the Debye mass is rather different from that in ref.\cite{chandra1}. Here, we have determined the Debye mass
from the chromo-electric  permittivity recently obtained by Ranjan and Ravishankar\cite{akranjan}. On the other hand, the method adopted in ref.\cite{chandra1} is 
based on the earlier works of Kelly {\it et. al}\cite{cm1,cm4}. We remark that the two approaches are equivalent at the one loop level provided that the phenomenological charge is fixed appropriately.

For the interacting QGP 
the Debye mass can be written by, 
\begin{equation}
\label{dm}
M^2_D=8\pi^3 Q^2 T^2\frac{N_c}{3} \bigg( g_2^{\prime}(z_g)+\frac{N_f}{2N_c} f^{\prime}_2(z_g)\bigg).
\end{equation} 
To see how the interactions influence the Debye screening, we consider the ratio, $R_{M_{D}}=(M_D)^2/(M^{I}_D)^2$,
which is given as,
\begin{equation}
R_{M_D}=\frac{2N_c g^{\prime}_2(z_g)+N_f f^{\prime}_2(z_q)}{2 N_c +N_f}.
\end{equation}
The expression of $R_{M_D}$ is the same as that for relative susceptibility ${\mathcal R}^{\prime}$ which follows from the 
expression for the Debye mass of full QCD in terms of the response function ( eq.\ref{eqs}). It is clear from  figs. 4 and 5
that the inclusion of the interactions significantly lower the Debye mass. The Debye mass will approach its 
ideal value only asymptotically. 
We have shown the behavior of the screening length as a function of temperature for EOS1 and EOS$\delta$ in fig. 7 and fig.8 respectively.
To compare these results with lattice predictions of Zantow and Kazmarek\cite{zantow}, we choose
$T_c=0.27$ GeV for pure gauge theory and $T_c=0.203, 0.197$ for 2- and 3-flavor QCD respectively. 
The screening length for both EOS1 and EOS2 show qualitative agreement with these results in \cite{zantow}.
As expected, the agreement is more pronounced for EOS$\delta$ as compare to EOS1.
We shall now proceed to determine the dissociation temperatures for various 
quarkonia states in a hot QCD medium.
\begin{figure}[htb]
\label{fig7}
\vspace*{-78mm}
\hspace*{-43mm}
\psfig{figure=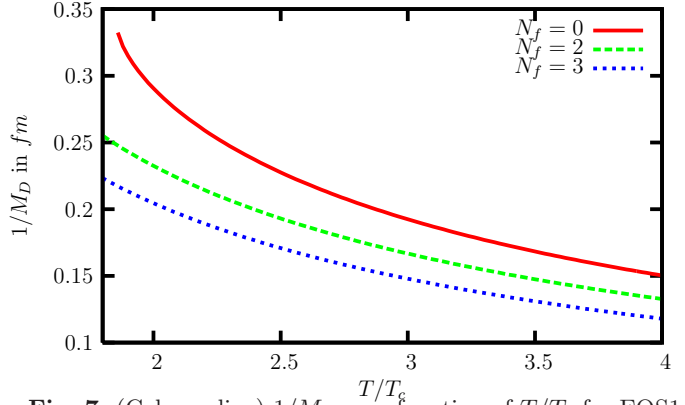,width=150mm}
\vspace*{-90mm}
\caption{(Color online) $1/M_D$ as a function of $T/T_c$ for EOS1 }
\end{figure}
\begin{figure}[htb]
\label{fig8}
\vspace*{-80mm}
\hspace*{-43mm}
\psfig{figure=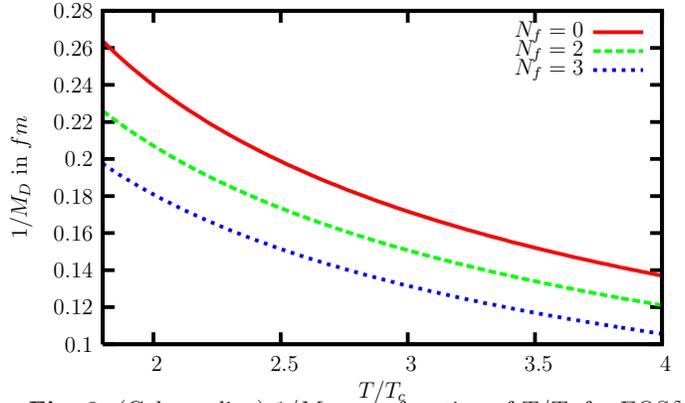,width=150mm}
\vspace*{-90mm}
\caption{(Color online) $1/M_D$ as a function of $T/T_c$ for EOS$\delta$.}
\end{figure}

\subsection{Dissociation temperatures}
To determine the dissociation temperatures of heavy quarkonia, we follow a well known criterion--whenever the {\it rms} radius, $r_{q{\bar q}}\ge 1/M_D$  for a particular heavy quarkonia bound state, the 
state will dissociate in the medium. The equality sign will yield the dissociation temperature. To this end, we shall employ the ${\it rms}$ radius of various bound states of charmonium and bottomonium in ref.\cite{satz}. We have shown the dissociation temperatures for various quarkonia in Table I along with the flavor dependence. The dissociation temperatures decrease with increasing number of flavors. The  $J/\Psi$ dissociation temperature varies between $1.9 T_c$ to $1.5 T_c$ for EOS$\delta$ and $2.28 T_c$ to $1.58 T_c$ for EOS1. Note that we have determined the dissociation temperatures by 
assuming the validity of  EOS1 and EOS$\delta$ temperatures as small as  $T\ge 1.5 T_c$.

 \begin{table}
\label{table1}
\caption{The dissociation temperature($T_D$) for various quarkonia states (in unit of $T_c$).
Note that we have employed one loop expression for the running coupling constant\cite{shaung} and
$T_c=$0.27, 0.203 and 0.197 for $N_f=0$, $N_f=2$ and $N_f=3$ flavor QCD \cite{zantow} respectively.}
\centering
\begin{tabular}{|l|l|l|l|l|}
\hline
Hot EOS& Quarkonium &$N_f=0$& $N_f=2$&$N_f=3$\\
&&&&\\
\hline\hline
EOS$\delta$ &$J/\Psi$&1.92&1.60&1.50\\
&$\Psi^\prime$&$<$ 1.50&$<$ 1.50&$<$ 1.50\\
&$\Upsilon$&3.88&3.30&2.77\\
&$\Upsilon^\prime$&1.70&1.50&$<$ 1.50\\
&$\chi_c$&$<$ 1.50&$<$ 1.50&$<$ 1.50\\
&$\chi_b$&2.21&1.87&1.60\\
\hline
EOS1 &$J/\Psi$&2.28&1.85&1.58\\
&$\Psi^\prime$&$<$1.50&$<$ 1.50&$<$ 1.50\\
&$\Upsilon$&4.35&3.74&3.22\\
&$\Upsilon^\prime$&2.05&1.64& 1.50\\
&$\chi_c$&$<$ 1.50&$<$ 1.50&$<$ 1.50\\
&$\chi_b$&2.60&2.15&1.84\\
\hline
\end{tabular}
\end{table}

We now  turn our attention to compare hot QCD estimates for dissociation temperatures
with other theoretical works. In a recent paper, Satz\cite{satz} has studied the dissociation of quarkonia states by studying their in-medium behavior. These estimates were based on the Schr\"odinger equation for the Cornell potential. In a more recent work,  Alberico {\it et al}\cite{prd75} reported the dissociation temperatures for charmonium and bottomonium states for $N_f=0$ and $N_f=2$
QCD.  They have solved the  Schr\"odinger equation for the charmonium and bottomonium states at finite temperature in the presence of a temperature dependent potential-- computed from the lattice QCD. 
The estimates for $T_D$ shown in Table I are consistent with these estimates\cite{satz,prd75} and also 
consistent with  other lattice results\cite{sdata} for quenched QCD and predictions  of dynamical $N_f=2$ QCD by Aarts {\it et al}\cite{gert}. Along these results, we wish to mention the very recent estimates on dissociation temperature reported by M\'ocsy and P\'etreeczky\cite{moscky} and Agotiya, Chandra and Patra\cite{chandra4}. Their estimates for $J/\Psi$ dissociation temperatures 
and for $\Upsilon$ are significantly smaller than the earlier results and results quoted in Table 1.

\section{Conclusions and Outlook}
In conclusion, we have successfully extracted the quasi-free particle content of two hot QCD equations of states and used them  to determine the
chromo-electric permittivities within the standard Boltzmann-Vlasov kinetic approach. The Abelian and the
non-Abelian components of the permittivity are obtained, for pure gauge theory and the full QCD. We have shown that the effect of the interactions is to merely renormalize the magnitude of the effective color charge, $Q$. We have determined the Debye mass from the chromo-electric susceptibility which yields the 
leading order HTL result for an ideal QGP. The viability of the two EOS, especially EOS$\delta$ is thus phenomenologically well supported. 
It would also be of interest to extend the analysis to other signatures like strangeness enhancement, and also for QGP with a finite baryonic chemical potential \cite{avrn,ipp} and the HTL and HDL equations of state\cite{rebh,rebh1}. 

\vspace{6mm}
\noindent {\bf Acknowledgments:}
VC thanks the Raman Research Institute, Bangalore (India) for hospitality where a part of this work was
completed. VC also acknowledges C.S.I.R., New Delhi (India) for the financial support.

%%%%%%%%%%%%%%%%%%%%%%%%%%%%%%%%%%%%%%%%%%%%%%%%%%%%%%%%%%%%%%%%%%%%%%%%%%%

\end{document}